\documentclass[twoside,twocolumn,9pt]{article}
\usepackage{extsizes}
\usepackage[super,sort&compress,comma]{natbib} 
\usepackage[version=3]{mhchem}
\usepackage[left=1.5cm, right=1.5cm, top=1.785cm, bottom=2.0cm]{geometry}
\usepackage{balance}
\usepackage{mathptmx}
\usepackage{sectsty}
\usepackage{graphicx} 
\usepackage{lastpage}
\usepackage[format=plain,justification=justified,singlelinecheck=false,font={stretch=1.125,small,sf},labelfont=bf,labelsep=space]{caption}
\usepackage{float}
\usepackage{fancyhdr}
\usepackage{fnpos}
\usepackage[english]{babel}
\addto{\captionsenglish}{%
  
}
\usepackage{array}
\usepackage{droidsans}
\usepackage{charter}
\usepackage[T1]{fontenc}
\usepackage[usenames,dvipsnames]{xcolor}
\usepackage{setspace}
\usepackage[compact]{titlesec}
\usepackage{hyperref}

\usepackage{epstopdf}

\definecolor{cream}{RGB}{222,217,201}
\usepackage{cuted}
\begin{document}

\pagestyle{fancy}
\thispagestyle{plain}
\fancypagestyle{plain}{
\renewcommand{\headrulewidth}{0pt}
}

\makeFNbottom
\makeatletter
\renewcommand\LARGE{\@setfontsize\LARGE{15pt}{17}}
\renewcommand\Large{\@setfontsize\Large{12pt}{14}}
\renewcommand\large{\@setfontsize\large{10pt}{12}}
\renewcommand\footnotesize{\@setfontsize\footnotesize{7pt}{10}}
\makeatother

\renewcommand{\thefootnote}{\fnsymbol{footnote}}
\renewcommand\footnoterule{\vspace*{1pt}%
\color{cream}\hrule width 3.5in height 0.4pt \color{black}\vspace*{5pt}} 
\setcounter{secnumdepth}{5}

\makeatletter 
\renewcommand\@biblabel[1]{#1}            
\renewcommand\@makefntext[1]%
{\noindent\makebox[0pt][r]{\@thefnmark\,}#1}
\makeatother 
\renewcommand{\figurename}{\small{Fig.}~}
\sectionfont{\sffamily\Large}
\subsectionfont{\normalsize}
\subsubsectionfont{\bf}
\setstretch{1.125} 
\setlength{\skip\footins}{0.8cm}
\setlength{\footnotesep}{0.25cm}
\setlength{\jot}{10pt}
\titlespacing*{\section}{0pt}{4pt}{4pt}
\titlespacing*{\subsection}{0pt}{15pt}{1pt}

\fancyfoot{}
\fancyfoot[LO,RE]{\vspace{-7.1pt}\includegraphics[height=9pt]{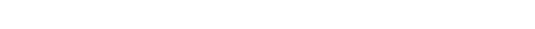}}
\fancyfoot[CO]{\vspace{-7.1pt}\hspace{13.2cm}\includegraphics{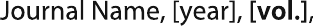}}
\fancyfoot[CE]{\vspace{-7.2pt}\hspace{-14.2cm}\includegraphics{head_foot/RF}}
\fancyfoot[RO]{\footnotesize{\sffamily{1--\pageref{LastPage} ~\textbar  \hspace{2pt}\thepage}}}
\fancyfoot[LE]{\footnotesize{\sffamily{\thepage~\textbar\hspace{3.45cm} 1--\pageref{LastPage}}}}
\fancyhead{}
\renewcommand{\headrulewidth}{0pt} 
\renewcommand{\footrulewidth}{0pt}
\setlength{\arrayrulewidth}{1pt}
\setlength{\columnsep}{6.5mm}
\setlength\bibsep{1pt}

\makeatletter 
\newlength{\figrulesep} 
\setlength{\figrulesep}{0.5\textfloatsep} 

\newcommand{\topfigrule}{\vspace*{-1pt}%
\noindent{\color{cream}\rule[-\figrulesep]{\columnwidth}{1.5pt}} }

\newcommand{\botfigrule}{\vspace*{-2pt}%
\noindent{\color{cream}\rule[\figrulesep]{\columnwidth}{1.5pt}} }

\newcommand{\dblfigrule}{\vspace*{-1pt}%
\noindent{\color{cream}\rule[-\figrulesep]{\textwidth}{1.5pt}} }

\makeatother

\twocolumn[
  \begin{@twocolumnfalse}
{\includegraphics[height=30pt]{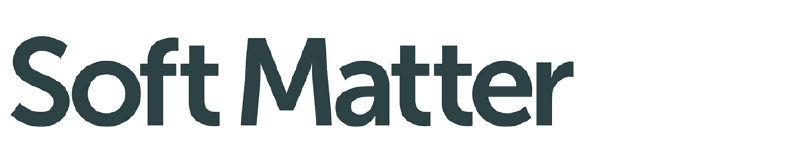}\hfill\raisebox{0pt}[0pt][0pt]{\includegraphics[height=55pt]{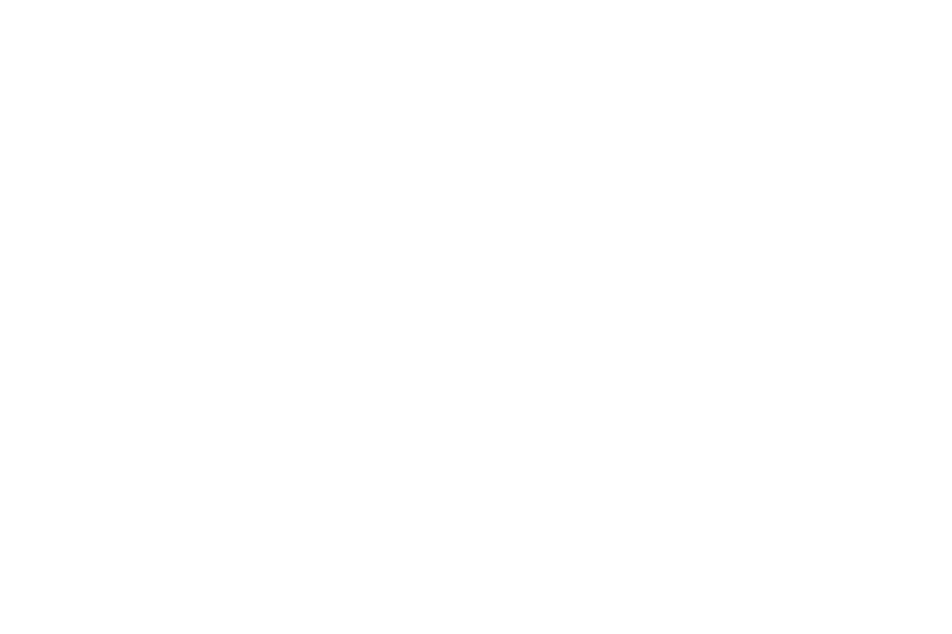}}\\[1ex]
\includegraphics[width=18.5cm]{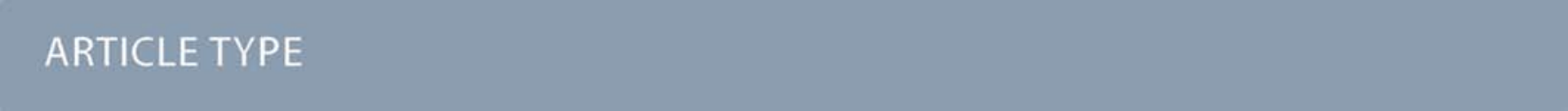}}\par
\vspace{1em}
\sffamily
\begin{tabular}{m{4.5cm} p{13.5cm} }

\includegraphics{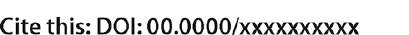} & \noindent\LARGE{\textbf{Analytical prediction of logarithmic Rayleigh scattering in amorphous solids from tensorial heterogeneous elasticity with power-law disorder}} \\
\vspace{0.3cm} & \vspace{0.3cm} \\

 & \noindent\large{Bingyu Cui\textit{$^{a}$} and Alessio Zaccone$^{\ast}$\textit{$^{a,b,c\ddag}$}} \\

\includegraphics{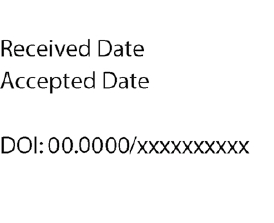} & \noindent\normalsize{The damping or attenuation coefficient of sound waves in solids due to impurities scales with the wavevector to the fourth power, also known as Rayleigh scattering. In amorphous solids, Rayleigh scattering may be enhanced by a logarithmic factor although computer simulations offer conflicting conclusions regarding this enhancement and its microscopic origin. We present a tensorial replica field-theoretic derivation based on heterogeneous or fluctuating elasticity (HE), which shows that long-range (power-law) spatial correlations of the elastic constants, is the origin of the logarithmic enhancement to Rayleigh scattering of phonons in amorphous solids. We also consider the case of zero spatial fluctuations in the elastic constants, and of power-law decaying fluctuations in the internal stresses. Also in this case the logarithmic enhancement to the Rayleigh scattering law can be derived from the proposed tensorial HE framework.} \\

\end{tabular}

 \end{@twocolumnfalse} \vspace{0.6cm}

  ]

\renewcommand*\rmdefault{bch}\normalfont\upshape
\rmfamily
\section*{}
\vspace{-1cm}


\footnotetext{\textit{$^{a}$~Cavendish Laboratory, University of Cambridge, JJ Thomson Avenue, CB3 0HE Cambridge, U.K.; E mail: alessio.zaccone@unimi.it}}
\footnotetext{\textit{$^{b}$~Department of Physics ``A. Pontremoli", University of Milan, via Celoria 16, 20133 Milano, Italy }}
\footnotetext{\textit{$^{c}$~Statistical Physics Group, Department of Chemical
Engineering and Biotechnology, University of Cambridge, Philippa Fawcett Drive, CB3 0AS Cambridge, U.K.} }

\footnotetext{~Electronic Supplementary Information (ESI) available: [details of any supplementary information available should be included here]. See DOI: 10.1039/cXsm00000x/}

\footnotetext{\ddag~Additional footnotes to the title and authors can be included \textit{e.g.}\ `Present address:' or `These authors contributed equally to this work' as above using the symbols: \ddag, \textsection, and \P. Please place the appropriate symbol next to the author's name and include a \texttt{\textbackslash footnotetext} entry in the the correct place in the list.}


\section{Introduction}
Amorphous solids exhibit anomalous thermal and vibrational properties at low temperature. Thanks to improved scattering experiments, as well as numerical simulations, recent years have witnessed important achievements in our understanding of glassy materials. One interesting property, as reported in \cite{Gelin2016}, is that long-wavelength phonons are more strongly attenuated in glasses than in ordinary crystalline solids, with an attenuation coefficient that scales with wavenumber $k$ as $\sim -k^{d+1}\ln{k}$ (in dimension $d$), thus with a logarithmic enhancement compared with the well known Rayleigh scattering law $\sim k^{d+1}$, whose validity has never been questioned in the last fifty years of studies of sound attenuation in amorphous materials \cite{Strutt1871}.

To be more specific, a compilation of many experiments with X-ray and light scattering demonstrates that the wavenumber dependence of the longitudinal sound attenuation coefficient, $\Gamma_L(k)$ is in general divided into three regimes~\cite{Carini,Baldi,Ruffle,Masciovecchio,Giordano,Ruta,Monaco2011,Monaco2016,Monaco2009}: (1) $\Gamma_L(k)\sim k^2$ for low $k$; (2) $\Gamma_L(k)\sim k^4$ for an intermediate $k$ regime; and (3) $\Gamma_L(k)\sim k^2$ for large $k$. It has been proven that the $k^2$ to $k^4$ transition for sound attenuation in a large frequency regime is mainly harmonic. In contrast, the $k^2$ dependence in the low-frequency regime was already clearly related to viscous attenuation caused by anharmonicity. Most computer studies address the sound attenuation problem at zero temperature in order to remove anharmonic effects and thus isolate the effect of disorder. In particular, regardless of system size, a recent numerical study of 2D systems reveals that the logarithmic correction to the cubic scaling, $\Gamma_\lambda(k)\sim -k^3\ln k$ ($\lambda=L,T$ stands for longitudinal and transverse) emerges in the boson peak (BP) regime, while it disappears as the wavenumber approaches the continuum limit, where $\Gamma_\lambda(k)\sim k^3$ is recovered \cite{Mizuno2018}. Authors in \cite{Gelin2016} even revisited data in experimental systems to confirm the attenuation coefficient indeed corresponds to the enhanced $-k^{d+1}\ln k$ law.

To rationalise the observed logarithmic correction to the Rayleigh law, one interpretation is to invoke the existence of correlated inhomogeneities of the elastic constants within the framework of fluctuating or heterogeneous elasticity (HE), yet neither quantitative nor qualitative arguments have been presented \cite{Marruzzo2013,Mizuno2013,Mizuno2014,Marruzzo2013Vibrational,Mizuno2019Impact,Wang2019}. Also, the possible relation between the logarithmic correction to the Rayleigh law and the long-range nature of elastic modulus has been questioned in \cite{Mizuno2018,Moriel2019}. In particular, in \cite{Moriel2019} simulations results indicate that the log enhancement to Rayleigh scattering does not correlate with fluctuations in the elastic constants, but appears, instead, to be strongly correlated with spatially heterogeneous internal stresses. Finally, a recent analysis in Ref.~\cite{Lemaitre2019} argues that HE is unable to predict the logarithmic enhancement.

In this paper, by developing a fully tensorial replica field theory for athermal amorphous systems with power-law decay in elastic constant correlations (or in stress fields), we reveal the origin of the enhanced phonon attenuation, especially where the logarithmic enhancement is prompted. The analytical theory shows that the logarithmic enhancement is either due to the long-range power-law correlations of elastic constants \cite{John1983a, John1983b} or (as shown in the Appendix B) to long-range power-law correlations of the internal stresses (with no fluctuations in the elastic constants)~\cite{Jie2020,Maier2018}, which is the key ingredient in our framework leading to the prediction of the logarithmic enhancement. Some previous works dealing with mean-field theory confirm the Rayleigh scattering law without the logarithmic factor. In those works, there is no power-law decay in correlations of elasticity\cite{Maurer2004, Shimada2019,Wyart2010,Kohler2013,DeGiuli2014}. We will mainly consider systems with similar elastic properties as in \cite{Gelin2016}, in an athermal regime where scattering is purely harmonic (no viscous/anharmonic dissipation involved). However, such systems usually have coupled internal longitudinal and transverse propagators: the explicit form of damping is thus not as clearly defined \cite{lifshitz1986}. Hence, the present work demonstrates that, contrary to claims of Ref.~\cite{Lemaitre2019}, heterogeneous elasticity (HE) in the fully tensorial formulation developed here for the first time is indeed able to recover the anomalous Rayleigh scattering observed in simulations. In our analysis we will work essentially within the linear acoustic dispersion relation regime.

\section{Formalism}
Throughout this paper, we focus on 2D systems. All results can be generalized to 3D case, by letting $\alpha,\beta,\kappa,\chi=x,y,z$ go through full Cartesian components and specifying bond orientation through the pair of angles $\phi,\theta$: $\underline{n}_{ij}=(\cos\phi_{ij}\sin\theta_{ij},\sin\phi_{ij}\sin\theta_{ij},\cos\theta_{ij})$. In elastic media, deformations of a generic material point (or a particle) are expressed in terms of microscopic displacements $\underline{u}$, defined as the current position of the particle at time $t$, $\underline{r}(\underline{\mathring{r}},t)$ minus its initially position located at $\underline{\mathring{r}}$, i.e. $\underline{u}=\underline{r}(\underline{\mathring{r}},t)-\underline{\mathring{r}}$. In the absence of body forces and assuming spatially uniform density $\rho$, the Lagrangian form of the elastic wave equation can be written as \cite{Gelin2016}
\begin{equation}
\rho\frac{\partial^2u^\alpha(\underline{\mathring{r}})}{\partial t^2}=\frac{\partial}{\partial\mathring{r}^\beta}\left[S^{\alpha\beta\kappa\chi}(\underline{\mathring{r}})\frac{\partial u^\kappa(\underline{\mathring{r}})}{\partial\mathring{r}^\chi}\right]
\end{equation}
with
\begin{equation}
S^{\alpha\beta\kappa\chi}(\underline{\mathring{r}})=C^{\alpha\beta\kappa\chi}(\underline{\mathring{r}})+\delta^{\alpha\kappa}\sigma^{\beta\chi}(\underline{\mathring{r}})
\end{equation}
where $C^{\alpha\beta\kappa\chi}$ and $\sigma^{\beta\chi}$ are the elastic constants and the Cauchy stress in the reference configuration, respectively. Greek subscripts refer to Cartesian coordinates and $\delta^{\alpha\kappa}$ denotes the Kronecker delta.
We note that, with the pair interaction $V_{ij}$, $C_{ij}^{\alpha\beta\kappa\chi}=h_{ij}n_{ij}^\alpha n_{ij}^\beta n_{ij}^\kappa n_{ij}^\chi$ where $r_{ij}$ is the interatomic distance, $\underline{n}_{ij}$ is the unit vector pointing from $i$ to $j$ and $h_{ij}=V_{ij}^{\prime\prime}(r_{ij})r_{ij}^2-V^\prime_{ij}(r_{ij})r_{ij}$ \cite{Lemaitre2006}. Prime denotes the derivative with respect to distance. The second term on RHS in Eq. (2) involves the pair contributions to the internal stress field and hence carries long-range spatial correlations due to stress. Following the assumptions of Ref.\cite{Gelin2016}, we ignore the contribution of spatial correlations in stress tensors. The influence of long-range fluctuations in stress tensor on elastic waves will be studied and discussed in Appendices A \& B for the case where, instead, no fluctuations in the elastic moduli exist.

Writing $\underline{n}_{ij}=(\cos\theta_{ij},\sin\theta_{ij})$, the elastic constants appear to be ofthe form $C_{ij}^{\alpha\beta\kappa\chi}=h_{ij}\cos^n\theta_{ij}\sin^{4-n}\theta_{ij}, n=0,...,4$. There are, hence, five local constants for each pair, they are~\cite{Gelin2016}
\begin{align}
C_{ij}^1&=h_{ij};\quad
C_{ij}^2=h_{ij}\cos(2\theta_{ij}),\notag\\
C_{ij}^3&=h_{ij}\sin(2\theta_{ij});\quad
C_{ij}^4=h_{ij}\cos(4\theta_{ij}),\notag\\
C_{ij}^5&=h_{ij}\sin(4\theta_{ij}).
\end{align}
Contributions of each pair to the Lam$\acute{e}$ constants are $\mu_{ij}=(1/8)(C_{ij}^1-C_{ij}^4)$ and $\lambda_{ij}=(1/8)(C_{ij}^1+C_{ij}^4)$.
We are able to express effective elastic constants $S^{\alpha\beta\kappa\chi}\approx C^{\alpha\beta\kappa\chi}$ in terms of these five local constants:
\begin{align}
C_{ij}^{xxxx}&=\frac{C_{ij}^4}{8}+\frac{C_{ij}^2}{2}+\frac{3C_{ij}^1}{8}\notag\\
C_{ij}^{xxxy}&=C_{ij}^{xxyx}=C_{ij}^{xyxx}=C_{ij}^{yxxx}=\frac{C_{ij}^5}{8}+\frac{C_{ij}^3}{4}\notag\\
C_{ij}^{xxyy}&=C_{ij}^{xyxy}=C_{ij}^{yxxy}=C_{ij}^{xyyx}=C_{ij}^{yxyx}=C_{ij}^{yyxx}=\frac{C_{ij}^1}{8}-\frac{C_{ij}^4}{8}\notag\\
C_{ij}^{yyyx}&=C_{ij}^{yyxy}=C_{ij}^{yxyy}=C_{ij}^{xyyy}=-\frac{C_{ij}^5}{8}+\frac{C_{ij}^3}{4}\notag\\
C_{ij}^{yyyy}&=\frac{C_{ij}^4}{8}-\frac{C_{ij}^2}{2}+\frac{3C_{ij}^1}{8}
\end{align}

Then, Eq. (1) becomes
\begin{strip}
\begin{align}
\rho\frac{\partial^2 u^x(\underline{\mathring{r}})}{\partial t^2}&=\frac{\partial}{\partial\mathring{r}^x}\left[\left(\frac{C^4}{8}+\frac{C^2}{2}+\frac{3C^1}{8}\right)\frac{\partial u^x}{\partial\mathring{r}^x}+\left(\frac{C^5}{8}+\frac{C^3}{4}\right)\left(\frac{\partial u^x}{\partial\mathring{r}^y}+\frac{\partial u^y}{\partial\mathring{r}^x}\right)+\left(\frac{C^1}{8}-\frac{C^4}{8}\right)\frac{\partial u^y}{\partial\mathring{r}^y}\right]\notag\\
&+\frac{\partial}{\partial\mathring{r}^y}\left[\left(\frac{C^5}{8}+\frac{C^3}{4}\right)\frac{\partial u^x}{\partial\mathring{r}^x}+\left(\frac{C^1}{8}-\frac{C^4}{8}\right)\left(\frac{\partial u^x}{\partial\mathring{r}^y}+\frac{\partial u^y}{\partial\mathring{r}^x}\right)+\left(-\frac{C^5}{8}+\frac{C^3}{4}\right)\frac{\partial u^y}{\partial\mathring{r}^y}\right]\notag\\
\rho\frac{\partial^2 u^y(\underline{\mathring{r}})}{\partial t^2}&=\frac{\partial}{\partial\mathring{r}^x}\left[\left(\frac{C^5}{8}+\frac{C^3}{4}\right)\frac{\partial u^x}{\partial\mathring{r}^x}+\left(\frac{C^1}{8}-\frac{C^4}{8}\right)\left(\frac{\partial u^x}{\partial\mathring{r}^y}+\frac{\partial u^y}{\partial\mathring{r}^x}\right)+\left(-\frac{C^5}{8}+\frac{C^3}{4}\right)\frac{\partial u^y}{\partial\mathring{r}^y}\right]\notag\\
&+\frac{\partial}{\partial\mathring{r}^y}\left[\left(\frac{C^1}{8}-\frac{C^4}{8}\right)\frac{\partial u^x}{\partial\mathring{r}^x}+\left(-\frac{C^5}{8}+\frac{C^3}{4}\right)\left(\frac{\partial u^x}{\partial\mathring{r}^y}+\frac{\partial u^y}{\partial\mathring{r}^x}\right)+\left(\frac{C^4}{8}-\frac{C^2}{2}+\frac{3C^1}{8}\right)\frac{\partial u^y}{\partial\mathring{r}^y}\right].
\end{align}
\end{strip}
\subsection{Toy model with vanishing $C^1, C^2, C^4, C^5$ and long-range correlation in $C^3$}
To probe the simplest possible scenario of long-range correlations in the elastic constants, we assume $C^1(\underline{r}),C^2(\underline{r}),C^4(\underline{r}),C^5(\underline{r})=0$ while $C^3(\underline{r})\equiv C(\underline{r})=\rho C_0+\rho\tilde{C}(\underline{r})$ is expressed in terms of its mean value plus a random part, i.e. $\overline{\tilde{C}(\underline{r})}=0$ and $\overline{\tilde{C}(\underline{r}')\tilde{C}(\underline{r}'+\underline{r})}=B(\underline{r})=\gamma\cos(4\theta)/(r^2+a^2)\equiv\cos(4\theta)B(r)$ for some constants $\gamma$ and $a$, where the final form is in polar coordinates. In principle, $a$ might also depend on $\underline{r}$ as long as it decays faster than $\sim r^2$ when $r\rightarrow\infty$. Here, we just let it be a constant. In other words, only the effect of non-vanishing $C^3$ is considered, whose spatial autocorrelation scales as $1/r^2$. The power-law decay in the self-correlation of elasticity, $B(r)$, has been numerically investigated in simulations in \cite{Gelin2016}. Similar behaviour in the spatial correlation of mass was analysed in detail within a scalar model of wave propagation in \cite{John1983b}.

Equation (5) then reduces to
\begin{strip}
\begin{align}
\rho\frac{\partial^2 u^x(\underline{\mathring{r}})}{\partial t^2}&=\frac{1}{4}\left[\frac{\partial C(\underline{\mathring{r}})}{\partial\mathring{r}^x}\frac{\partial u^x}{\partial\mathring{r}^y}+\frac{\partial C(\underline{\mathring{r}})}{\partial\mathring{r}^y}\frac{\partial u^x}{\partial\mathring{r}^x}+2C(\underline{\mathring{r}})\frac{\partial^2 u^x}{\partial\mathring{r}^x\partial\mathring{r}^y}+\frac{\partial C(\underline{\mathring{r}})}{\partial\mathring{r}^x}\frac{\partial u^y}{\partial\mathring{r}^x}+\frac{\partial
C(\underline{\mathring{r}})}{\partial\mathring{r}^y}\frac{\partial u^y}{\partial\mathring{r}^y}+C(\underline{\mathring{r}})\left(\frac{\partial^2u^y}{\partial\mathring{r}^{x}\partial\mathring{r}^{x}}+\frac{\partial^2u^y}{\partial\mathring{r}^{y}\partial\mathring{r}^{y}}\right)\right]\notag\\
&=\frac{1}{4}\sum_{\alpha\neq \beta}\left[\nabla_\alpha(C\nabla_\beta)\right]u^x+\frac{1}{4}\sum_\alpha\left[\nabla_\alpha(C\nabla_\alpha)\right]u^y\notag\\
\rho\frac{\partial^2 u^y(\underline{\mathring{r}})}{\partial t^2}&=\frac{1}{4}\left[\frac{\partial C(\underline{\mathring{r}})}{\partial\mathring{r}^x}\frac{\partial u^y}{\partial\mathring{r}^y}+\frac{\partial C(\underline{\mathring{r}})}{\partial\mathring{r}^y}\frac{\partial u^y}{\partial\mathring{r}^x}+2C(\underline{\mathring{r}})\frac{\partial^2 u^y}{\partial\mathring{r}^x\partial\mathring{r}^y}+\frac{\partial C(\underline{\mathring{r}})}{\partial\mathring{r}^x}\frac{\partial u^x}{\partial\mathring{r}^x}+\frac{\partial
C(\underline{\mathring{r}})}{\partial\mathring{r}^y}\frac{\partial u^x}{\partial\mathring{r}^y}+C(\underline{\mathring{r}})\left(\frac{\partial^2u^x}{\partial\mathring{r}^{x}\partial\mathring{r}^{x}}+\frac{\partial^2u^x}{\partial\mathring{r}^{y}\partial\mathring{r}^{y}}\right)\right]\notag\\
&=\frac{1}{4}\sum_\alpha\left[\nabla_\alpha(C\nabla_\alpha)\right]u^x+\frac{1}{4}\sum_{\alpha\neq \beta}\left[\nabla_\alpha(C\nabla_\beta)\right]u^y.
\end{align}
\end{strip}

In frequency space, upon letting $z^2=\omega^2+i0$, the equation of motion of the frequency-dependent displacement vector $\underline{u}(\underline{r},z)$ is (we have dropped the ring)
\begin{align}
&A(z)\underline{u}(\underline{r},z)=0,\notag\\
\textup{with}\qquad &A^{xx}=A^{yy}=-\rho z^2-\frac{1}{4}\sum_{\alpha\neq\beta}\left[\nabla_\alpha(C\nabla_\beta)\right], \notag\\
&A^{xy}=A^{yx}=\frac{1}{4}\sum_\alpha\left[\nabla_\alpha(C\nabla_\alpha)\right].
\end{align}
The spatial correlation of $C(\underline{r})$ may be implemented by the probability distribution for its fluctuating part,
\begin{equation}
P[\tilde{C}(\underline{r})]=P_0\exp\left[-\frac{1}{2}\int d^2rd^2r'\tilde{C}(\underline{r})B^{-1}(\underline{r}-\underline{r}')\tilde{C}(\underline{r}')\right]
\end{equation}
where $B^{-1}$ is the inverse of $B(\underline{r}-\underline{r}')$ such that
\begin{equation}
\int d^2pB(\underline{r}-\underline{p})B^{-1}(\underline{p}-\underline{q})=\delta(\underline{r}-\underline{q}),
\end{equation}
while $P_0$ is a normalization factor. The Lagrangian is expressed as (rescaled by $\rho$),
\begin{strip}
\begin{align}
L&=\frac{1}{2}\int d^2r\underline{u}^TA\underline{u}=\frac{1}{2}\int d^2r\left[u^x(A_{xx}u^x+A_{xy}u^y)+u^y(A_{yx}u^x+A_{yy}u^y)\right]\notag\\
&=\frac{1}{2}\int d^2r\{-z^2\underline{u}\cdot\underline{u}-\frac{1}{4}\sum_{\alpha\neq\beta}u^x\left[\nabla_\alpha(C\nabla_\beta u^x)\right]-\frac{1}{4}\sum_\alpha u^x\left[\nabla_\alpha(C\nabla_\alpha u^y)\right]-\frac{1}{4}\sum_{\alpha\neq \beta}u^y\left[\nabla_\alpha(C\nabla_\beta u^y)\right]-\frac{1}{4}\sum_\alpha u^y\left[\nabla_\alpha(C\nabla_\alpha u^x)\right]\}\notag\\
&=\frac{1}{2}\int d^2r\{-z^2\underline{u}\cdot\underline{u}-\frac{1}{4}\sum_{\alpha\neq\beta}\left[\nabla_\alpha[u^xC\nabla_\beta u^x]-(\nabla_\alpha u^x)C(\nabla_\beta u^x)\right]-\frac{1}{4}\sum_\alpha\left[\nabla_\alpha[u^xC\nabla_\alpha u^y]-C(\nabla_\alpha u^x)(\nabla_\alpha u^y)\right]-x\longleftrightarrow y\}\notag\\
&=\frac{1}{2}\int d^2r\{-z^2\underline{u}\cdot\underline{u}-\frac{1}{4}\nabla_x[u^xC\nabla_yu^x+u^yC\nabla_yu^y+u^xC\nabla_xu^y+u^yC\nabla_xu^x]-\frac{1}{4}\nabla_y[x\longleftrightarrow y]\notag\\
&+\frac{1}{2}C[(\nabla_xu^x)(\nabla_yu^x)+(\nabla_xu^y)(\nabla_yu^y)+(\nabla_xu^x)(\nabla_xu^y)+(\nabla_yu^x)(\nabla_yu^y)]\}\notag\\
&=\frac{1}{2}\int d^2r\{-z^2\underline{u}\cdot\underline{u}-\frac{1}{4}\nabla_x[...]-\frac{1}{4}\nabla_y[...]+\frac{1}{2}C(\nabla\cdot\underline{u})(\nabla_xu^y+\nabla_yu^x)\}\notag\\
&=\frac{1}{2}\int d^2r\{-z^2\underline{u}\cdot\underline{u}+\frac{1}{2}C(\nabla\cdot\underline{u})(\nabla_xu^y+\nabla_yu^x)\}
\end{align}
\end{strip}
where the object $[...]$ vanishes on the boundary.
Using the replica-field representation, the generating functional for calculating the averaged Green's function takes the form
\begin{strip}
\begin{align}
&\langle Z(0)\rangle\equiv\lim_{n\rightarrow0}\int\mathcal{D}[\underline{u}_a(\underline{r})]\mathcal{D}[\tilde{C}(\underline{r})]P_0\notag\\
&\times\exp\left[-\frac{1}{2}\sum_{a=1}^n\int d^2r\{-z^2\underline{u}_a(\underline{r})^2+\frac{C(\underline{r})}{2}(\nabla\cdot\underline{u}_a(\underline{r}))(\nabla_xu^y_a+\nabla_yu^x_a)\}-\frac{1}{2}\int d^2rd^2r'\tilde{C}(\underline{r})B^{-1}(\underline{r}-\underline{r}')\tilde{C}(\underline{r}')\right]\notag\\
&=\lim_{n\rightarrow0}\int\mathcal{D}[\underline{u}_a(\underline{r})]\mathcal{D}[\tilde{C}(\underline{r})]P_0\exp\left[-\frac{1}{2}\sum_{a=1}^n\int d^2r\{{-z^2\underline{u}_a(\underline{r})^2+\frac{1}{2}C_0(\nabla\cdot\underline{u}_a(\underline{r}))(\nabla_xu^y_a+\nabla_yu^x_a)+\frac{1}{2}\tilde{C}(\underline{r})(\nabla\cdot\underline{u}_a(\underline{r}))(\nabla_xu^y_a+\nabla_yu^x_a)\}}\right.\notag\\
&\qquad\left.{-\frac{1}{2}\int d^2rd^2r'\tilde{C}(\underline{r})B^{-1}(\underline{r}-\underline{r}')\tilde{C}(\underline{r}')}\right]\notag\\
&\approx\lim_{n\rightarrow0}\int\mathcal{D}[\underline{u}_a(\underline{r})]\exp\left[{-\frac{1}{2}\sum_{a=1}^n\int d^2r\left\{-z^2\underline{u}_a(\underline{r})^2+\frac{1}{2}C_0(\nabla\cdot\underline{u}_a(\underline{r}))(\nabla_xu^y_a+\nabla_yu^x_a)\right\}}\right.\notag\\
&\qquad\left.{+\frac{1}{32}\sum_{a,b=1}^n\int d^2rd^2r'(\nabla\cdot\underline{u}_a(\underline{r}))(\nabla_xu^y_a(\underline{r})+\nabla_yu^x_a(\underline{r}))B(\underline{r}-\underline{r}')(\nabla\cdot\underline{u}_b(\underline{r}'))(\nabla_xu^y_b(\underline{r}')+\nabla_yu^x_b(\underline{r}'))}\right]
\end{align}
\end{strip}
where $a=1,...,n$ is a replica index (same as $b$), and the $n\rightarrow0$ limit eliminates the determinant factor. By means of a Hubbard-Stratonovich transformation, we introduce the effective matrix fields $\Lambda^{\alpha\beta\kappa\chi}_{ab}(\underline{r},\underline{r}',z)$ to replace the $\tilde{C}(\underline{r})$ in the harmonic part of the effective equation of motion. Then $\langle Z(0)\rangle$ becomes
\begin{strip}
\begin{align}
\langle Z(0)\rangle&\approx\lim_{n\rightarrow0}\int\mathcal{D}[\underline{u}_a(\underline{r})]\mathcal{D}[\Lambda_{ab}^{\alpha\beta\kappa\chi}(\underline{r},\underline{r}',z)]\Lambda_0\notag\\
&\times\exp\left\{{-\frac{1}{2}\sum_{a=1}^n\int d^2r\left[-z^2\underline{u}_a(\underline{r})^2-\frac{C_0}{4}\left(\sum_{\alpha\neq\beta}(u^x_a\nabla_\alpha\nabla_\beta u^x_a+u^y_a\nabla_\alpha\nabla_\beta u^y_a)+\sum_{\alpha}(u^x_a\nabla_\alpha\nabla_\alpha u^y_a+u^y_a\nabla_\alpha\nabla_\alpha u^x_a)\right)\right]}\right.\notag\\
&\qquad\left.{-\frac{1}{2}\sum_{a,b=1}^n\sum_{\alpha\kappa\chi\beta=x,y}\int d^2rd^2r'\left[\Lambda_{ab}^{\alpha\kappa\chi\beta}(\underline{r},\underline{r}',z)B^{-1}(\underline{r}-\underline{r}')\sum_{\kappa'\chi'}\Lambda_{ab}^{\alpha\beta\kappa'\chi'}(\underline{r},\underline{r}',z)
-\frac{1}{4}u^\alpha_a(\underline{r})\nabla_{\kappa}\Lambda_{ab}^{\alpha\beta\kappa\chi}(\underline{r},\underline{r}',z)\nabla_{\chi}u^\beta_b(\underline{r}')\right]}\right\}
\end{align}
\end{strip}
where $\Lambda_{ab}^{\alpha\beta\kappa\chi}=0$ if $\alpha=\beta,\kappa=\chi$ or $\alpha\neq\beta,\kappa\neq\chi$.  The way $\Lambda_{ab}^{\alpha\beta\kappa\chi}$ is introduced is to make Eq. (12) consistent with Eq. (7). We will see in the following calculations that the way to index $\Lambda^{\alpha\beta\kappa\chi}$ will be fulfilled by $\epsilon^{\alpha\beta\kappa\chi}$. The normalization constant is represented as $\Lambda_0$.
The generating function including source $J_{ab}^{\alpha\beta}(\underline{r},\underline{r}')$ is
\begin{strip}
\begin{subequations}
\begin{align}
\langle Z(J)\rangle&=\lim_{n\rightarrow0}\int\mathcal{D}[\underline{u}_a(\underline{r})]\mathcal{D}[\Lambda_{ab}^{\alpha\beta\kappa\chi}(\underline{r},\underline{r}',z)]\Lambda_0\notag\\
&\times\exp\left\{{-\frac{1}{2}\sum_{a=1}^n\int d^2r\left[-z^2\underline{u}_a(\underline{r})^2-\frac{C_0}{4}\left(\sum_{\alpha\neq\beta}(u^x_a\nabla_\alpha\nabla_\beta u^x_a+u^y_a\nabla_\alpha\nabla_\beta u^y_a)+\sum_{\alpha}(u^x_a\nabla_\alpha\nabla_\alpha u^y_a+u^y_a\nabla_\alpha\nabla_\alpha u^x_a)\right)\right]}\right.\notag\\
&\qquad\left.{-\frac{1}{2}\sum_{a,b=1}^n\sum_{\alpha\kappa\chi\beta=x,y}\int d^2rd^2r'\sum_{\kappa'\chi'}\Lambda_{ab}^{\alpha\kappa\chi\beta}B^{-1}(\underline{r}-\underline{r}')\Lambda_{ab}^{\alpha\kappa'\chi'\beta}
-\frac{1}{4}u^\alpha_a(\underline{r})\nabla_\kappa\Lambda_{ab}^{\alpha\beta\kappa\chi}}\nabla_\chi u^\beta_b(\underline{r}')+2J_{ab}^{\alpha\beta}\Lambda_{ab}^{\alpha\beta\kappa\chi}\right\}\notag\\
&=\lim_{n\rightarrow0}\int\mathcal{D}[\underline{u}_a(\underline{r})]\mathcal{D}[\Lambda_{ab}^{\alpha\beta\kappa\chi}(\underline{r},\underline{r}',z)]\Lambda_0\notag\\
&\times\exp\left\{-\frac{1}{2}\sum_{a,b=1}^n\sum_{\alpha\kappa\chi\beta=x,y}\int d^2rd^2r'\left[\underline{u}_a(\underline{r})A_{ab}(\Lambda_{ab}^{\alpha\kappa\chi\beta})\underline{u}_b(\underline{r}')
+\sum_{\kappa'\chi'}\Lambda_{ab}^{\alpha\beta\kappa\chi}B^{-1}(\underline{r}-\underline{r}')\Lambda_{ab}^{\alpha\beta\kappa'\chi'}
+2J_{ab}^{\alpha\beta}\Lambda_{ab}^{\alpha\beta\kappa\chi}\right]\right\}
\end{align}
\begin{flalign}
&\text{where}&\notag
\end{flalign}
\begin{align}
A_{ab}(\Lambda)&\equiv\delta(\underline{r}-\underline{r}')\delta^{ab}\left(\begin{array}{cc}
-z^2-\frac{C_0}{4}\sum_{\alpha\neq\beta}\nabla_\alpha\nabla_\beta & -\frac{C_0}{4}\sum_\alpha\nabla_\alpha\nabla_\alpha\\
-\frac{C_0}{4}\sum_\alpha\nabla_\alpha\nabla_\alpha & -z^2-\frac{C_0}{4}\sum_{\alpha\neq\alpha}\nabla_\alpha\nabla_\alpha
\end{array}\right)\notag\\
&-\frac{1}{4}\sum_{\kappa\chi}\left(\begin{array}{cc}
\nabla_\kappa\Lambda^{xx\kappa\chi}_{ab}(\underline{r},\underline{r}',z)\nabla_\chi & \nabla_\kappa\Lambda^{xy\kappa\chi}_{ab}(\underline{r},\underline{r}',z)\nabla_\chi\\
\nabla_\kappa\Lambda^{yx\kappa\chi}_{ab}(\underline{r},\underline{r}',z)\nabla_\chi & \nabla_\kappa\Lambda^{yy\kappa\chi}_{ab}(\underline{r},\underline{r}',z)\nabla_\chi
\end{array}\right).
\end{align}
\end{subequations}
\end{strip}
By evaluating derivatives of $\langle Z(J)\rangle$ with respect to $J^{\alpha\beta}_{ab}$ at $J^{\alpha\beta}_{ab}=0$, we are able to find the averaged Green's function of $\Lambda_{ab}^{\alpha\beta\kappa\chi}$.
Integrating $\underline{u}_a$ out in Eq. (12), we obtain a field theory involving only the $\Lambda$ field:
\begin{strip}
\begin{align}
&\langle Z(0)\rangle\approx\lim_{n\rightarrow0}\int\mathcal{D}[\Lambda]\exp\left\{-\frac{1}{2}\sum_{a,b=1}^n\sum_{\alpha,\beta,\kappa,\chi}\left(\ln\det A(\Lambda^{\alpha\beta\kappa\chi}_{ab})+\sum_{\kappa'\chi'}\int d^2rd^2r'\Lambda^{\alpha\beta\kappa\chi}B^{-1}(\underline{r}-\underline{r}')\Lambda^{\alpha\beta\kappa'\chi'}\right)\right\}.
\end{align}
\end{strip}
We seek a saddle-point such that the exponential in Eq. (14) is stationary, which corresponds to the mean-field theory of spatially correlated disorder of the coherent-potential approximation (CPA) for the one-particle Green's function. A saddle point $\Lambda'$ of the $\Lambda$ field is a point such that the exponential in Eq. (14) contains no terms linear in a small fluctuation $\hat{\Lambda}\equiv\Lambda-\Lambda'$. On the other hand, if we expand the Lagrangian about $\Lambda'$, keeping only quadratic displacements in $\hat{\Lambda}$, then the saddle-point value of $\Lambda$ determines the averaged one-particle Green's function:
\begin{equation}
\Lambda'=\langle\Lambda\rangle
\end{equation}
Expanding in $\hat{\Lambda}$, $A_{ab}(\Lambda)$ is written as
\begin{align}
&A_{ab}(\Lambda)=A_{ab}(\Lambda')\notag\\
&-\frac{1}{4}\sum_{\kappa\chi}\left(\begin{array}{cc}
\nabla_\kappa\hat{\Lambda}^{xx\kappa\chi}_{ab}(\underline{r},\underline{r}',z)\nabla_\chi & \nabla_\kappa\hat{\Lambda}^{xy\kappa\chi}_{ab}(\underline{r},\underline{r}',z)\nabla_\chi\\
\nabla_\kappa\hat{\Lambda}^{yx\kappa\chi}_{ab}(\underline{r},\underline{r}',z)\nabla_\chi & \nabla_\kappa\hat{\Lambda}^{yy\kappa\chi}_{ab}(\underline{r},\underline{r}',z)\nabla_\chi
\end{array}\right)\notag\\
&\equiv A_{ab}(\Lambda')+\hat{A}_{ab}(\hat{\Lambda}).
\end{align}
Making use of the identity
\begin{align}
\ln\det(A(\Lambda')+\hat{A}(\hat{\Lambda}))&=\ln\det(A(\Lambda'))\notag\\
&+\sum_n^\infty\frac{(-1)^{n+1}}{n}\text{Tr}(\underbrace{A^{-1}\hat{A}...A^{-1}\hat{A}}_{2n}),
\end{align}
the corresponding saddle-point equations can be solved with a replica-diagonal field $\Lambda_{ab}^{\alpha\beta\kappa\chi}(\underline{r},\underline{r}',z)=\Sigma^{\alpha\beta}(\underline{r},\underline{r}',z)\delta^{ab}$, which is relevant to the evaluation of the averaged one-particle Green's function. Non-diagonal saddle points need only be taken into account at the stage of renormalization \cite{Mckane1981}. Then we have $\ln\det(A)=\text{Tr}\ln(A)$. The self-energy $\langle\Sigma^{\alpha\beta}\rangle$ measuring the average response of the $\alpha$th component of the displacement field at $\underline{r}$ to an impulse in the $\beta$th component at $\underline{r}'$, can be determined by taking
\begin{align}
&\frac{\delta}{\delta\Sigma^{\alpha\beta}}\left(\text{Tr}\ln A(\Sigma^{\alpha\beta})+\int d^2rd^2r'\Sigma^{\alpha\beta} B^{-1}(\underline{r}-\underline{r}')\Sigma^{\alpha\beta}\right)\notag\\
&=0
\end{align}
at $\Sigma_0^{\alpha\beta}$, which yields
\begin{subequations}
\begin{align}
\langle\Sigma_0^{\alpha\beta}\rangle=\frac{1}{8}\sum_{\kappa\chi}\nabla_\kappa B(\underline{r}-\underline{r}')\nabla_\chi \langle G_0(r^\alpha,r'^\beta,z)\rangle\epsilon^{\alpha\beta\kappa\chi}
\end{align}
\begin{align}
&G_0(r^x,r'^x,z)=\left[-z^2-\frac{(C_0+\Sigma^{xx}_0)}{4}(\nabla_x\nabla_y+\nabla_y\nabla_x)\right]^{-1}\\
&G_0(r^y,r'^y,z)=\left[-z^2-\frac{(C_0+\Sigma^{yy}_0)}{4}(\nabla_x\nabla_y+\nabla_y\nabla_x)\right]^{-1}\\
&G_0(r^x,r'^y)=\left[-\frac{(C_0+\Sigma^{xy}_0)}{4}(\nabla_x\nabla_x+\nabla_y\nabla_y)\right]^{-1}\\
&G_0(r^y,r'^x)=\left[-\frac{(C_0+\Sigma^{yx}_0)}{4}(\nabla_x\nabla_x+\nabla_y\nabla_y)\right]^{-1}
\end{align}
\end{subequations}
where $\epsilon^{\alpha\beta\kappa\chi}=0$ if $\alpha=\beta,\kappa=\chi$ or $\alpha\neq\beta, \kappa\neq\chi$. Translational invariance holds after taking the ensemble average, hence the CPA Green's function $G_0$ depends only on the difference between two points in space.

\subsection{Theory with non-zero $C^1, C^2, C^4,C^5$ and long-range correlations in $C^3$}
\begin{figure}
\centering
\includegraphics[width=0.45\textwidth]{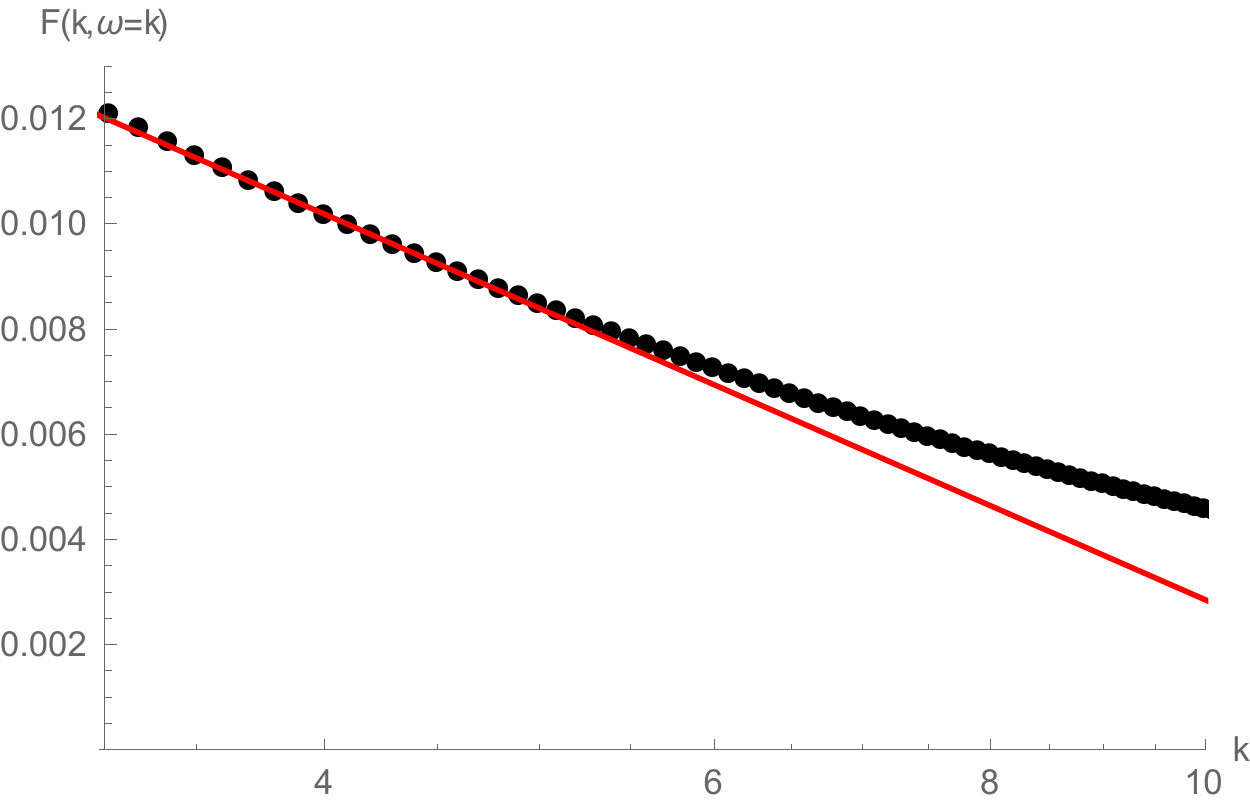}
\caption{Fitting of $F(k,\omega)$ (i.e. the numerical integral in Eq. (26)) (symbols) with logarithmic function $-p_0\ln(p_1k)$ (solid line). Parameters are $a=10, c=1, p_0=0.008$ and $p_1=0.07$.}
\label{fig:logfit}
\end{figure}

We weaken our condition on the other elastic constants by letting $C^1, C^2, C^4$ and $C^5$ be all non-zero constants. The propagator $A$ in $A(z)\underline{u}(\underline{r},z)=0$ takes the form (scaled with $\rho$)
\begin{align}
A^{\alpha\beta}=-z^2\delta^{\alpha\beta}-\sum_{\kappa\chi}\mathcal{C}^{\alpha\beta\kappa\chi}\nabla_\kappa\nabla_\chi
-\frac{1}{4}\sum_{\kappa\chi}[\nabla_\kappa(\tilde{C}\nabla_\chi)]\epsilon^{\alpha\beta\kappa\chi}
\end{align}
where $\mathcal{C}^{\alpha\beta\kappa\chi}$ corresponds to the $\underline{r}$-independent part of elastic constants $C^i,i=1,2,3,4,5$. The explicit form is not important and we do not provide it here.
The Lagrangian becomes
\begin{align}
L&=\frac{1}{2}\int d^2r\{-z^2u^2+\frac{1}{2}\tilde{C}(\nabla\cdot\underline{u})(\nabla_xu^y+\nabla_yu^x)\notag\\
&-\sum_{\alpha\beta\kappa\chi}u^\alpha \mathcal{C}^{\alpha\beta\kappa\chi}\nabla_\kappa\nabla_\chi u^\beta)\}
\end{align}
and $\langle Z(J)\rangle$ is
\begin{strip}
\begin{subequations}
\begin{align}
\langle Z(J)\rangle
&=\lim_{n\rightarrow0}\int\mathcal{D}[\underline{u}^a(\underline{r})]\mathcal{D}[\Lambda_{ab}^{\alpha\beta\kappa\chi}(\underline{r},\underline{r}',z)]\Lambda_0\notag\\
&\times\exp\left\{-\frac{1}{2}\sum_{a,b=1}^n\sum_{\alpha\beta\kappa\chi=x,y}\int d^2rd^2r'\left[u_a^\alpha(\underline{r})A_{ab}(\Lambda_{ab}^{\alpha\beta\kappa\chi})u_b^\beta(\underline{r}')
+\sum_{\kappa'\chi'}\Lambda_{ab}^{\alpha\beta\kappa\chi}B^{-1}(\underline{r}-\underline{r}')\Lambda_{ab}^{\alpha\beta\kappa'\chi'}
+2J_{ab}^{\alpha\beta}\Lambda_{ab}^{\alpha\beta\kappa\chi}\right]\right\}
\end{align}
\begin{flalign}
&\text{with}&\notag
\end{flalign}
\begin{align}
A_{ab}^{\alpha\beta\kappa\chi}(\Lambda)&\equiv\delta(\underline{r}-\underline{r}')\delta^{ab}\left(-z^2\delta^{\alpha\beta}
-\sum_{\kappa\chi}\mathcal{C}^{\alpha\beta\kappa\chi}\nabla_\kappa\nabla_\chi\right)
-\frac{1}{4}\sum_{\kappa\chi}\nabla_\kappa\Lambda_{ab}^{\alpha\beta\kappa\chi}(\underline{r},\underline{r'},z)\nabla_\chi
\end{align}
\end{subequations}
\end{strip}
Again, letting $\Lambda_{ab}^{\alpha\beta\kappa\chi}=\Sigma^{\alpha\beta}\delta^{ab}$ and finding the saddle point of $\text{Tr}\ln A(\Sigma^{\alpha\beta})+\int d^2rd^2r'\Sigma^{\alpha\beta}B(\underline{r}-\underline{r}')\Sigma^{\alpha\beta}$, we obtain the self-consistent equations for the self-energy and the Green's function:
\begin{strip}
\begin{subequations}
\begin{align}
\langle\Sigma_0^{\alpha\beta}\rangle&=\frac{1}{8}\sum_{\kappa\chi}\nabla_\kappa B(\underline{r}-\underline{r}')\nabla_\chi \langle G_0(r^\alpha,r'^\beta,z)\rangle\epsilon^{\alpha\beta\kappa\chi}\\
G_0(r^x,r'^x,z)&=\left[-z^2-\left(\frac{C^4}{8}+\frac{C^2}{2}+\frac{3C^1}{8}\right)\nabla_x\nabla_x-\left(\frac{C^1}{8}-\frac{C^4}{8}\right)\nabla_y\nabla_y
-\left(\frac{C^5}{8}+\frac{(C_0+\Sigma^{xx}_0)}{4}\right)\left(\nabla_x\nabla_y+\nabla_y\nabla_x\right)\right]^{-1}\\
G_0(r^y,r'^y,z)&=\left[-z^2-\left(\frac{C^4}{8}-\frac{C^2}{2}+\frac{3C^1}{8}\right)\nabla_y\nabla_y-\left(\frac{C^1}{8}-\frac{C^4}{8}\right)\nabla_x\nabla_x
-\left(-\frac{C^5}{8}+\frac{(C_0+\Sigma^{yy}_0)}{4}\right)\left(\nabla_x\nabla_y+\nabla_y\nabla_x\right)\right]^{-1}\\
G_0(r^x,r'^y)&=\left[-\left(\frac{C^1}{8}-\frac{C^4}{8}\right)(\nabla_x\nabla_y+\nabla_y\nabla_x)-\frac{(C_0+\Sigma^{xy}_0)}{4}(\nabla_x\nabla_x+\nabla_y\nabla_y)
-\frac{C^5}{8}(\nabla_x\nabla_x-\nabla_y\nabla_y)\right]^{-1}\\
G_0(r^y,r'^x)&=\left[-\left(\frac{C^1}{8}-\frac{C^4}{8}\right)(\nabla_x\nabla_y+\nabla_y\nabla_x)-\frac{(C_0+\Sigma^{yx}_0)}{4}(\nabla_x\nabla_x+\nabla_y\nabla_y)
-\frac{C^5}{8}(\nabla_x\nabla_x-\nabla_y\nabla_y)\right]^{-1}.
\end{align}
\end{subequations}
\end{strip}
Defining the Fourier transform as
\begin{equation}
\Sigma(\underline{k},z)\equiv\int d^2(\underline{r}-\underline{r}')e^{i\underline{k}(\underline{r}-\underline{r}')}\Sigma(\underline{r}-\underline{r}',z),
\end{equation}
the condition on the one-particle CPA Green's function may be rewritten in momentum space:
\begin{strip}
\begin{subequations}
\begin{align}
&\langle\Sigma_0^{\alpha\beta}\rangle=-\frac{1}{4}\sum_{\kappa\chi}\epsilon^{\alpha\beta\kappa\chi}k_\kappa k_\chi\int d^2q\tilde{B}(\underline{k}-\underline{q})\langle G_0(\underline{q})\rangle;\qquad
\tilde{B}(\underline{k})\equiv\int d^2re^{i\underline{k}\cdot\underline{r}}B(\underline{r})\\
&G_0(k^{x},k^x,z)=\left[-z^2+\left(\frac{C^4}{8}+\frac{C^2}{2}+\frac{3C^1}{8}\right)k^xk^x+\left(\frac{C^1}{8}-\frac{C^4}{8}\right)k^yk^y+\left(\frac{C^5}{4}+\frac{C_0+\Sigma^{xx}_0}{2}\right)k^xk^y\right]^{-1}\\
&G_0(k^{y},k^y,z)=\left[-z^2+\left(\frac{C^4}{8}-\frac{C^2}{2}+\frac{3C^1}{8}\right)k^yk^y+\left(\frac{C^1}{8}-\frac{C^4}{8}\right)k^xk^x+\left(-\frac{C^5}{4}+\frac{C_0+\Sigma^{yy}_0}{2}\right)k^xk^y\right]^{-1}\\
&G_0(k^{x},k^y)=\left[\left(\frac{C^1}{4}-\frac{C^4}{4}\right)k^xk^y+\frac{(C_0+\Sigma^{xy}_0)}{4}k^2+\frac{C^5}{8}(k^xk^x-k^yk^y)\right]^{-1};\quad G_0(k^{y},k^x)=\left[\left(\frac{C^1}{4}-\frac{C^4}{4}\right)k^xk^y+\frac{(C_0+\Sigma^{yx}_0)}{4}k^2+\frac{C^5}{8}(k^xk^x-k^yk^y)\right]^{-1}
\end{align}
\end{subequations}
\end{strip}
which must be solved self-consistently since the self-energy of the Green's function involves the full propagator itself (this self-consistency has been ignored in Ref.~\cite{Lemaitre2019}). In the weak scattering limit, approximate solutions are possible because Im$\Sigma(\underline{k},z)$ is small compared with $C_0$ and also the imaginary part of the propagator takes the form of a $\delta$-function, Im$\langle G_0(\underline{q})\rangle\propto\delta(\omega^2-q^2)$ upon averaging over all possible directions of dummy variable $\underline{q}$ and upon re-scaling redundant constants. Upon taking the imaginary part, the correlation function in Eq.(25a) can be evaluated using the $\delta$-function and we calculate the self-energy to be
\begin{align}
&\langle\Sigma_0^{\alpha\beta}\rangle\propto\sum_{\kappa\chi}\epsilon^{\alpha\beta\kappa\chi}k_\kappa k_\chi\int_0^\infty\frac{rJ_4(kr)J_0(\omega r)}{r^2+a^2}dr,
\label{Bessel}
\end{align}
where $J_0$ and $J_4$ are (modified) Bessel functions. The detailed derivation is outlined in Appendix C. Making use of a linear dispersion relation with constant speed of sound, numerical computation reveals that the integral $F(k,\omega)=\int_0^\infty [rJ_4(kr)J_0(\omega r)/(r^2+a^2)]dr\sim-\ln k$ across a broad range, from low to intermediate values, of $k$. Figure \ref{fig:logfit} shows one typical plot for such fitting. In other words, the logarithmic dependence is caused by the integral of $r/(r^{2} + a^{2})$, while the Bessel functions in the integrand are responsible for bending the overall shape of $F(k,\omega)$ away from the log asymptote, and thus for restricing the logarithmic dependence to an intermediate range of $k$. This consideration is a further demonstration that the logarithmic correction stems from the power-law decay of correlations encoded in the integrand factor $r/(r^{2} + a^{2})$.
Thus, we obtain the averaged self-energy (susceptibility), in an intermediate range of $k$, as:
\begin{equation}
\text{Im}\langle\Sigma_0^{\alpha\beta}(k)\rangle\sim-k^2\ln k.
\end{equation}
where the linear dispersion relation $k\propto\omega$ is assumed.
A similar result was obtained by John and Stephen~\cite{John1983b} in a different context of Anderson localization of electromagnetic waves where a scalar model with power-law correlation in the spatially varying mass parameter was considered. To our knowledge, the one presented here is the first derivation of this effect in the context of phonon propagation in elastic media, thus accounting for the full tensorial nature of the problem.

We note that there are no purely longitudinal and transverse waves with respect to the direction of $\underline{k}$. This is different from the cases considered in \cite{Maurer2004,John1983a,John1983b}. However, cross terms (25d) essentially contribute nothing to the density of states. One might define a more general relation between damping and self-energy over different directions.
Hence Eq. (27) demonstrates that the self-energy of the phonon Green's function, which is closely related to the phonon attenuation coefficient, does indeed exhibit a logarithmic enhancement correction to the Rayleigh law as a result of power-law spatial correlations in, at least, the elastic constant $C^{3}$. Hence, this result holds for materials that are described within the heterogeneous elasticity framework.

\subsection{Theory with non-zero $C^1, C^4,C^5$ and long-range correlations in $C^2$ and $C^3$}
In addition to letting $C^3(\underline{r})\equiv\rho C_3+\rho\tilde{C}_3(\underline{r})$, we further require $C^2(\underline{r})\equiv\rho C_2+\rho\tilde{C}_2(\underline{r})$ with $\overline{\tilde{C}}_{2,3}(\underline{r})=0$ and $\overline{\tilde{C}_{2,3}(\underline{r}')\tilde{C}_{2,3}(\underline{r}'+\underline{r})}=B_{2,3}(\underline{r})=\gamma_{2,3}\cos(4\theta)/(r^2+a^2)$ for parameters $\gamma_{2,3}$, and parameter $a$.
In this case, the configurational average is due to spatial fluctuations of both $C^2$ and $C^3$ and is given by
\begin{equation}
P[\tilde{C}(\underline{r})]\propto\exp\left[-\frac{1}{2}\sum_{i=2,3}\int d^2rd^2r'\tilde{C}_i(\underline{r})B^{-1}_i(\underline{r}-\underline{r}')\tilde{C}_i(\underline{r}')\right].
\end{equation}
To implement the same formalism as above, we now introduce two effective fields to get the CPA for the one-particle Green's function.  The matrix operators (with effective fields) become
\newpage
\begin{strip}
\begin{align}
A^{\alpha\beta}&=-z^2\delta^{\alpha\beta}-\sum_{\kappa\chi}\mathcal{C}^{\alpha\beta\kappa\chi}\nabla_\kappa\nabla_\chi
-\frac{1}{4}\sum_{\kappa\chi}[\nabla_\kappa(\tilde{C}_3\nabla_\chi)]\epsilon^{\alpha\beta\kappa\chi}+\frac{1}{2}\nabla_\alpha\tilde{C}_2\nabla_\beta\delta^{\alpha\beta}\eta^{\alpha\beta}\\
A_{ab}^{\alpha\kappa\chi\beta}(\Lambda_2,\Lambda_3)&\equiv\delta(\underline{r}-\underline{r}')\left(-z^2\delta^{\alpha\beta}
-\sum_{\kappa\chi}\mathcal{C}^{\alpha\beta\kappa\chi}\nabla_\kappa\nabla\chi\right)
-\frac{1}{4}\sum_{\kappa\chi}\nabla_\kappa\Lambda_{ab,3}^{\alpha\beta\kappa\chi}(\underline{r},\underline{r'},z)\nabla_\chi\epsilon^{\alpha\beta\kappa\chi}
+\frac{1}{2}\nabla_\alpha\Lambda_{ab,2}^{\alpha\beta}\nabla_\beta\delta^{\alpha\beta}\eta^{\alpha\beta}
\end{align}
\end{strip}
where $\eta^{xx}=1$, $\eta^{yy}=-1$.
Repeating similar steps, the self-consistent equations take the following form
\begin{strip}
\begin{subequations}
\begin{align}
\langle\Sigma^{\alpha\beta}_{0,2}\rangle&=\frac{1}{4}\nabla_\alpha B_2(\underline{r}-\underline{r}')\nabla_\beta \langle G_0(r^\alpha,r'^\beta,z)\rangle\eta^{\alpha\beta};\qquad
\langle\Sigma^{\alpha\beta}_{0,3}\rangle=\frac{1}{8}\sum_{\kappa\chi}\nabla_\kappa B_3(\underline{r}-\underline{r}')\nabla_\chi \langle G_0(r^\alpha,r'^\beta,z)\rangle\epsilon^{\alpha\beta\kappa\chi}\\
G_0(r^x,r'^x,z)&=\left[-z^2-\left(\frac{C^4}{8}+\frac{C_2+\Sigma^{xx}_{0,2}}{2}+\frac{3C^1}{8}\right)\nabla_x\nabla_x-\left(\frac{C^1}{8}-\frac{C^4}{8}\right)\nabla_y\nabla_y
-\left(\frac{C^5}{8}+\frac{(C_3+\Sigma^{xx}_{0,3})}{4}\right)\left(\nabla_x\nabla_y+\nabla_y\nabla_x\right)\right]^{-1}\\
G_0(r^y,r'^y,z)&=\left[-z^2-\left(\frac{C^4}{8}-\frac{C_2-\Sigma^{yy}_{0,2}}{2}+\frac{3C^1}{8}\right)\nabla_y\nabla_y-\left(\frac{C^1}{8}-\frac{C^4}{8}\right)\nabla_x\nabla_x
-\left(-\frac{C^5}{8}+\frac{(C_3+\Sigma^{yy}_{0,3})}{4}\right)\left(\nabla_x\nabla_y+\nabla_y\nabla_x\right)\right]^{-1}\\
G_0(r^x,r'^y,z)&=\left[-\left(\frac{C^1}{8}-\frac{C^4}{8}\right)(\nabla_x\nabla_y+\nabla_y\nabla_x)-\frac{(C_3+\Sigma^{xy}_{0,3})}{4}(\nabla_x\nabla_x+\nabla_y\nabla_y)
-\frac{C^5}{8}(\nabla_x\nabla_x-\nabla_y\nabla_y)\right]^{-1}\\
G_0(r^y,r'^x,z)&=\left[-\left(\frac{C^1}{8}-\frac{C^4}{8}\right)(\nabla_x\nabla_y+\nabla_y\nabla_x)-\frac{(C_3+\Sigma^{yx}_{0,3})}{4}(\nabla_x\nabla_x+\nabla_y\nabla_y)
-\frac{C^5}{8}(\nabla_x\nabla_x-\nabla_y\nabla_y)\right]^{-1}.
\end{align}
\end{subequations}
\end{strip}
We note that, even if $B_2+B_3$ has no long-range tail, the net effect of the imaginary part of $\langle\Sigma^{\alpha\beta}_{0,2}\rangle+\langle\Sigma^{\alpha\beta}_{0,3}\rangle$ still exhibits log-enhancement. To see this more clearly, we write down $\Sigma_{0,2}^{xx}$ and $\Sigma_{0,3}^{xx}$,
\begin{align}
\langle\Sigma^{xx}_{0,2}\rangle&=\frac{1}{4}\nabla_x B_2(\underline{r}-\underline{r}')\nabla_x \langle G_0(r^x,r'^{x},z)\rangle\notag\\
\langle\Sigma^{xx}_{0,3}\rangle&=\frac{1}{8}\nabla_x B_3(\underline{r}-\underline{r}')\nabla_y \langle G_0(r^x,r'^x,z)\rangle\notag\\
&+\frac{1}{8}\nabla_y B_3(\underline{r}-\underline{r}')\nabla_x \langle G_0(r^x,r'^x,z)\rangle
\end{align}
Using the same arguments as in the last section, namely transforming to $\underline{k}$ space, we can easily verify that $\text{Im}\langle\Sigma_{0,2}\rangle\sim-k^2\ln k$ and $\text{Im}\langle\Sigma_{0,3}\rangle\sim-k^2\ln k$. Hence, we also have that $\text{Im}[\langle\Sigma_{0,2}^{xx}\rangle+\langle\Sigma_{0,3}^{xx}\rangle]\sim-k^2\ln k$ even if $B_2=-B_3$. Since this works the same for all components $\alpha, \beta$, we can conclude that
\begin{equation}
\text{Im}[\langle\Sigma_{0,2}^{\alpha\beta}\rangle+\langle\Sigma_{0,3}^{\alpha\beta}\rangle]\sim-k^2\ln k,
\end{equation}
which holds for of all components $\alpha,\beta$ of the self energy.

Hence, the logarithmic enhancement to Rayleigh scattering law remains confirmed also in the case of power-law spatial correlations in two elastic constants, $C^{2}$ and $C^{3}$.

This is the main result of this paper, which rigorously proves that power-law correlations lead to the logarithmic enhancement of Rayleigh scattering in amorphous solids, under the same conditions studied in numerical simulations in \cite{Gelin2016}, where this effect was observed. We note that, in Ref. \cite{Lemaitre2019}, it is reported that Rayleigh law without the logarithmic dependence is retrieved in the frame of fluctuating elasticity. The authors attribute this behavior to cancellation of the elastic correlations between spatial auto-correlations of non-diagonal part of local elastic coefficients. We emphasise that the imaginary part of self-energy obtained here in Eq. (33) does not split into purely transverse and longitudinal contributions, whereas that from Ref.~\cite{Lemaitre2019}'s method vanishes because those authors assume fully isotropic elasticity splitting into uncoupled longitudinal and transverse contributions (see e.g. Eq. (34) in the Supplementary Information of Ref. \cite{Lemaitre2019}), which does not correspond to the physical system in \cite{Gelin2016}, nor to other simulated systems where the effect was observed.

\section{Conclusion}
We have developed a fully tensorial replica field theory of heterogeneous elasticity which, in two dimensions, predicts that long-range elastic correlations cause a logarithmic enhancement to Rayleigh scattering of phonons in amorphous systems where internal stresses are absent. A similar calculation (reported in Appendix A and Appendix B) predicts the anomalous logarithmic correction to arise in the absence of fluctuations in the elastic constants, but in presence of power-law fluctuations in the internal stresses that have been recently discovered experimentally~\cite{Jie2020}. The mean-field method, which generally applies in infinite dimensions, might predict somewhat different results when it is compared to numerical results in finite dimension~\cite{Shimada2019}. In future work it may be worth studying how the logarithmic correction is affected by non-mean-field
effects within our framework of the HE, and possibly in synergy with numerical methods~\cite{Shimada2019}.

Recent work~\cite{Mizuno2018} on jammed harmonic sphere packings showed that the Rayleigh scattering law without the logarithmic correction is observed in the low wavevector limit in those systems, however the authors showed that in the systems they studied there were no long-range correlations in the elastic moduli. In ~\cite{Moriel2019}, the anomalous scattering was found to correlate with spatially fluctuating internal stresses in the absence of fluctuations of elastic constants.
Our analytical theory (in its formulation based on correlations in internal stresses), is able to provide a theoretical prediction for those observations.

We note that many numerical simulations addressing this problem so far only extract affine elastic constants, while the nonaffine contribution to elasticity~\cite{Lemaitre2006, Scossa-Romano, Zaccone2013,Mizuno2013} might be important in some systems and should be examined in detail in future work, although nonaffinity does not appear to be a necessary ingredient for the appearance of the anomalous logarithmic correction. Moreover, Ref.~\cite{Moriel2019} argued on the basis of numerical data of computer glasses that Rayleigh scaling is expected at low wavenumbers, where soft quasilocalized modes are scarce, while the logarithmically enhanced Rayleigh scaling of the form $\Gamma(k)\sim -k^{d+1}\ln k$ arises at higher $k$. We are aware that, in that case, the way to extract phononic and nonphononic excitations is different from \cite{Mizuno2018}, which might result in the different features of enhanced logarithmic dependence. We also note that the enhanced logarithmic Rayleigh scattering contributed by power-law elastic or stress correlations in our theory arises in a broad intermediate range of $k$. However, the discussion about this issue is beyond the scope of this paper, and will be studied in detail in future work.

Furthermore, our analysis is restricted to the athermal limit. At finite temperature, elastic correlators would receive additional effects from anharmonicity~\cite{Ruocco2019,Baggioli} and other thermal effects. We expect this problem to be important also for plasticity and yielding, which could be the object of future work.\\

\section*{Conflicts of interest}
There are no conflicts to declare.

\section*{Acknowledgements}
We are indebted to Hajime Tanaka for providing initial input, inspiration and motivation to perform this work and for hosting B.C. at the University of Tokyo. Useful discussions with E. Lerner, E. Bouchbinder, G. Ruocco, and E. M. Terentjev are gratefully acknowledged. This work was supported by the CSC-Cambridge Scholarship (B.C.) and by the US Army ARO Cooperative Agreement W911NF-19-2-0055 (A.Z.).\\


\begin{appendix}
\section{Equations of motion with stress correlations}
Here we instead assume that elastic constants $C^{\alpha\beta\kappa\chi}$ have no fluctuations, while fluctuations exist in the internal stresses, a situation encountered in glasses~\cite{Maier2018} and granular materials~\cite{Jie2020}. Writing $p_{ij}=-(1/2)V_{ij}^\prime(r_{ij})r_{ij}$, the local stress tensor can be decomposed at the pair level as:
\begin{align}
\sigma_{ij}^1&=p_{ij};\quad \sigma_{ij}^2&=-p_{ij}\cos(2\theta_{ij}),;\quad \sigma_{ij}^3&=-p_{ij}\sin(2\theta_{ij})
\end{align}
In this representation, $\sigma_{ij}^1$ is the pair-level pressure, while $\sigma_{ij}^2$ and $\sigma_{ij}^3=\sigma_{ij}^{xy}$ represent two shear stresses. We are able to express effective elastic constants $S^{\alpha\beta\kappa\chi}$ in this new representation. Given $\sigma^1,\sigma^2,\sigma^3$ and using Eq. (2) in the main text, we are able to obtain
\begin{align}
&S^{xxxx}(\underline{r})=C^{xxxx}+\sigma^2(\underline{r})-\sigma^1(\underline{r});\quad S^{xxxy}(\underline{r})=C^{xxxy}+\sigma^3(\underline{r})\notag\\
&S^{xxyx}(\underline{r})=C^{xxyx};\quad
S^{xxyy}(\underline{r})=C^{xxyy}\notag\\
&S^{xyxx}(\underline{r})=C^{xyxx}+\sigma^3(\underline{r});\quad
S^{xyxy}(\underline{r})=C^{xyxy}-\sigma^1(\underline{r})-\sigma^2(\underline{r})\notag\\
&S^{xyyx}(\underline{r})=C^{xyyx}\quad S^{xyyy}(\underline{r})=C^{xyyy}\notag\\
&S^{yxxx}(\underline{r})=C^{yxxx}\quad S^{yxxy}(\underline{r})=C^{yxxy}\notag\\
&S^{yxyx}(\underline{r})=C^{yxyx}+\sigma^2(\underline{r})-\sigma^1(\underline{r})\quad S^{yxyy}(\underline{r})=C^{yxyy}+\sigma^3(\underline{r})\notag\\
&S^{yyxx}(\underline{r})=C^{yyxx}\quad S^{yyxy}(\underline{r})=C^{yyxy}\notag\\
&S^{yyyx}(\underline{r})=C^{yxyy}+\sigma^3(\underline{r})\quad S^{yyyy}(\underline{r})=C^{yyyy}-\sigma^1(\underline{r})-\sigma^2(\underline{r}).
\label{newS}
\end{align}
Substituting Eq. (\ref{newS}) back to Eq. (1) gives (we drop the ring on $\underline{r}$):
\begin{strip}
\begin{align}
\rho\frac{\partial^2u^x}{\partial t^2}&=\frac{\partial}{\partial r^x}\left[\left(C^{xxxx}+\sigma^2(\underline{r})-\sigma^1(\underline{r})\right)\frac{\partial u^x}{\partial r^x}+\left(C^{xxxy}+\sigma^3(\underline{r})\right)\frac{\partial u^x}{\partial r^y}+C^{xxyx}\frac{\partial u^y}{\partial r^x}+C^{xxyy}\frac{\partial u^y}{\partial r^y}\right]\notag\\
&+\frac{\partial}{\partial r^y}\left[\left(C^{xyxx}+\sigma^3(\underline{r})\right)\frac{\partial u^x}{\partial r^x}+\left(C^{xyxy}-\sigma^1(\underline{r})-\sigma^2(\underline{r})\right)\frac{\partial u^x}{\partial r^y}+C^{xyyx}\frac{\partial u^y}{\partial r^x}+C^{xyyy}\frac{\partial u^y}{\partial r^y}\right]\notag\\
\rho\frac{\partial^2 u^y}{\partial t^2}&=\frac{\partial}{\partial r^x}\left[C^{yxxx}\frac{\partial u^x}{\partial r^x}+C^{yxxy}\frac{\partial u^x}{\partial r^y}+\left(C^{yxyx}+\sigma^2(\underline{r})-\sigma^1(\underline{r})\right)\frac{\partial u^y}{\partial r^x}+\left(C^{yxyy}+\sigma^3(\underline{r})\right)\frac{\partial u^y}{\partial r^y}\right]\notag\\
&+\frac{\partial}{\partial r^y}\left[C^{yyxx}\frac{\partial u^x}{\partial r^x}+C^{yyxy}\frac{\partial u^x}{\partial r^y}+\left(C^{yxyy}+\sigma^3(\underline{r})\right)\frac{\partial u^y}{\partial r^x}+\left(C^{yyyy}-\sigma^1(\underline{r})-\sigma^2(\underline{r})\right)\frac{\partial u^y}{\partial r^y}\right].
\end{align}
\end{strip}
\section{Prediction of logarithmic scattering with long-range decay in internal stress $\sigma^3$}
We assume that only $\sigma^3$ exhibits long-range behavior, i.e. $\sigma^3(\underline{r})=\rho\sigma_0+\rho\tilde{\sigma}(\underline{r})$ is expressed in terms of its mean value plus a random part, i.e. $\overline{\tilde{\sigma}(\underline{r})}=0$ and $\overline{\tilde{\sigma}(\underline{r}')\tilde{\sigma}(\underline{r}'+\underline{r})}=B(\underline{r})=\gamma\cos(4\theta)/(r^2+a^2)\equiv\cos(4\theta)B(r)$ for some constants $\gamma$ and $a$ again. All other elastic constants like $C^{\alpha\beta\kappa\chi}$ or $\sigma^1,\sigma^2$ are short-ranged and hence can be regarded as constant when $r$ is large. The long-range decay in shear stress correlations has been derived using generalized hydrodynamic theory in \cite{Maier2018}.
Then the elastic wave equation becomes
\begin{align}
\rho\frac{\partial^2 u^x(\underline{r})}{\partial t^2}&=\mathcal{S}^{x\beta\kappa\chi}\frac{\partial^2u^\kappa}{\partial r^\beta\partial r^\chi}+\frac{\partial\tilde{\sigma}(\underline{r})}{\partial r^x}\frac{\partial u^x}{\partial r^y}+\frac{\partial\tilde{\sigma}(\underline{r})}{\partial r^y}\frac{\partial u^x}{\partial r^x}\notag\\
\rho\frac{\partial^2 u^y(\underline{r})}{\partial t^2}&=\mathcal{S}^{y\beta\kappa\chi}\frac{\partial^2u^\kappa}{\partial r^\beta\partial r^\chi}+\frac{\partial\tilde{\sigma}(\underline{r})}{\partial r^x}\frac{\partial u^y}{\partial r^y}+\frac{\partial\tilde{\sigma}(\underline{r})}{\partial r^y}\frac{\partial u^y}{\partial r^x}
\end{align}
where $\mathcal{S}^{\alpha\beta\kappa\chi}$ corresponds to the $\underline{r}$-independent part of elastic or stress tensors.
Letting $z^2=\omega^2+i0$, the equation of motion of the frequency-dependent displacement vector $\underline{u}(\underline{r},z)$ is
\begin{align}
&A(z)\underline{u}(\underline{r},z)=0\quad\text{with}\\
&A^{\alpha\beta}=-z^2\delta^{\alpha\beta}
-\sum_{\kappa\chi}\mathcal{S}^{\alpha\beta\kappa\chi}\nabla_\kappa\nabla_\chi-\sum_{\kappa\neq\chi}(\nabla_\kappa[\tilde{\sigma}\nabla_\chi])\delta^{\alpha\beta}.
\end{align}
The fluctuation of $\sigma(\underline{r})$ is implemented by the probability distribution for its fluctuating part,
\begin{equation}
P[\tilde{\sigma}(\underline{r})]=P_0\exp\left[-\frac{1}{2}\int d^2rd^2r'\tilde{\sigma}(\underline{r})B^{-1}(\underline{r}-\underline{r}')\tilde{\sigma}(\underline{r}')\right].
\end{equation}
The Lagrangian is expressed as (scaled by $\rho$),
\begin{strip}
\begin{align}
&L=\frac{1}{2}\int d^2r\underline{u}^TA\underline{u}=u^x(A_{xx}u^x+A_{xy}u^y)+u^y(A_{yx}u^x+A_{yy}u^y)\notag\\
&=\frac{1}{2}\int d^2r\left\{-z^2\underline{u}\cdot\underline{u}-\sum_{\alpha\beta\kappa\chi}u^\alpha \mathcal{S}^{\alpha\beta\kappa\chi}\nabla_\kappa\nabla_\chi u^\beta-u^x\nabla_x(\tilde{\sigma}\nabla_y u^x)-u^x\nabla_y(\tilde{\sigma}\nabla_xu^x)-u^y\nabla_x(\tilde{\sigma}\nabla_yu^y)-u^y\nabla_y(\tilde{\sigma}\nabla_xu^y)\right\}\notag\\
&=\frac{1}{2}\int d^2r\left\{-z^2\underline{u}\cdot\underline{u}-\sum_{\alpha\beta\kappa\chi}u^\alpha \mathcal{S}^{\alpha\beta\kappa\chi}\nabla_\kappa\nabla_\chi u^\beta-[\nabla_x(u^x\tilde{\sigma}\nabla_yu^x)
+\nabla_y(u^x\tilde{\sigma}\nabla_xu^x)-2\tilde{\sigma}(\nabla_yu^x)(\nabla_xu^x)]-[x\leftrightarrow y]\right\}\notag\\
&=\frac{1}{2}\int d^2r\left\{-z^2\underline{u}\cdot\underline{u}-\sum_{\alpha\beta\kappa\chi}u^\alpha \mathcal{S}^{\alpha\beta\kappa\chi}\nabla_\kappa\nabla_\chi u^\beta+2\tilde{\sigma}[(\nabla_yu^y)(\nabla_xu^y)+(\nabla_xu^x)(\nabla_yu^x)]\right\}
\end{align}
\end{strip}
where the last equality holds because objects like $\nabla_x(u^x\tilde{\sigma}\nabla_yu^x)$ vanish on the boundary. Using the replica-field representation, the generating functional takes the form
\begin{strip}
\begin{align}
&\langle Z(0)\rangle\equiv
\lim_{n\rightarrow0}\int\mathcal{D}[\underline{u}_a(\underline{r})]\mathcal{D}[\tilde{\sigma}(\underline{r})]P_0\exp\left[-\frac{1}{2}\sum_{a=1}^n\int d^2r\{{-z^2u_a(\underline{r})^2+2\tilde{\sigma}[(\nabla_yu^y)(\nabla_xu^y)+(\nabla_xu^x)(\nabla_yu^x)]\}}\right.\notag\\
&\qquad\left.{-\sum_{\alpha\beta\kappa\chi}u^\alpha \mathcal{S}^{\alpha\beta\kappa\chi}\nabla_\kappa\nabla_\chi u^\beta-\frac{1}{2}\int d^2rd^2r'\tilde{\sigma}(\underline{r})B^{-1}(\underline{r}-\underline{r}')\tilde{\sigma}(\underline{r}')}\right]\approx\lim_{n\rightarrow0}\int\mathcal{D}[\underline{u}_a(\underline{r})]\exp\left[{-\frac{1}{2}\sum_{a=1}^n\int d^2r\left\{-z^2u_a(\underline{r})^2-\sum_{\alpha\beta\kappa\chi}u^\alpha \mathcal{S}^{\alpha\beta\kappa\chi}\nabla_\kappa\nabla_\chi u^\beta\right\}}\right.\notag\\
&\qquad\left.{+\frac{1}{2}\sum_{a,b=1}^n\int d^2rd^2r'[(\nabla_yu^y_a(\underline{r}))(\nabla_xu^y_a(\underline{r}))+(\nabla_xu^x_a(\underline{r}))(\nabla_yu^x_a(\underline{r}))]
B(\underline{r}-\underline{r}')
[(\nabla_yu^y_b(\underline{r}'))(\nabla_xu^y_b(\underline{r}'))+(\nabla_xu^x_b(\underline{r}'))(\nabla_yu^x_b(\underline{r}'))]}\right]
\end{align}
\end{strip}
where again $a,b=1,...,n$. We introduce effective matrix fields $\Lambda^{\alpha\beta\kappa\chi}_{ab}(\underline{r},\underline{r}',z)$ to replace the $\tilde{\sigma}(\underline{r})$ in the harmonic part of the effective equation of motion:
\begin{strip}
\begin{align}
\langle Z(0)\rangle&\approx\lim_{n\rightarrow0}\int\mathcal{D}[\underline{u}_a(\underline{r})]\mathcal{D}[\Lambda_{ab}^{\alpha\beta\kappa\chi}(\underline{r},\underline{r}',z)]\Lambda_0\cdot\exp\left\{{-\frac{1}{2}\sum_{a=1}^n\int d^2r\left[-z^2u_a(\underline{r})^2-\sum_{\alpha\beta\kappa\chi}u^\alpha \mathcal{S}^{\alpha\beta\kappa\chi}\nabla_\kappa\nabla_\chi u^\beta\right]}\right.\notag\\
&\qquad\left.{-\frac{1}{2}\sum_{a,b=1}^n\sum_{\alpha\beta,\kappa\neq\chi}\int d^2rd^2r'\left[\Lambda_{ab}^{\alpha\beta\kappa\chi}(\underline{r},\underline{r}',z)B^{-1}(\underline{r}-\underline{r}')\sum_{\kappa'\neq\chi'}\Lambda_{ab}^{\alpha\beta\kappa'\chi'}(\underline{r},\underline{r}',z)
-u^\alpha_a(\underline{r})\nabla_{\kappa}\Lambda_{ab}^{\alpha\beta\kappa\chi}(\underline{r},\underline{r}',z)\nabla_{\chi}u^\beta_b(\underline{r}')\delta^{\alpha\beta}\right]}\right\}.
\end{align}
\end{strip}
The way $\Lambda_{ab}^{\alpha\kappa\chi\beta}$ is introduced is to make Eq. (48) consistent with Eq. (42). $\Lambda_0$ is a normalization constant.
The generating function including source $J_{ab}^{\alpha\beta}(\underline{r},\underline{r}')$ is then:
\begin{strip}
\begin{subequations}
\begin{align}
\langle Z(J)\rangle&=\lim_{n\rightarrow0}\int\mathcal{D}[\underline{u}_a(\underline{r})]\mathcal{D}[\Lambda_{ab}^{\alpha\beta\kappa\chi}(\underline{r},\underline{r}',z)]\Lambda_0\notag\\
&\cdot\exp\left\{-\frac{1}{2}\sum_{a,b=1}^n\sum_{\alpha\beta,\kappa\neq\chi}\int d^2rd^2r'\left[\underline{u}_a(\underline{r})A_{ab}(\Lambda_{ab}^{\alpha\beta\kappa\chi})\underline{u}_b(\underline{r}')
+\sum_{\kappa'\neq\chi'}\Lambda_{ab}^{\alpha\beta\kappa\chi}B^{-1}(\underline{r}-\underline{r}')\Lambda_{ab}^{\alpha\beta\kappa'\chi'}
+2J_{ab}^{\alpha\beta}\Lambda_{ab}^{\alpha\beta\kappa\chi}\right]\right\}
\end{align}
\begin{flalign}
&\text{where}&\notag
\end{flalign}
\begin{align}
A_{ab}(\Lambda_{ab}^{\alpha\beta\kappa\chi})&\equiv\delta^{ab}\delta(\underline{r}-\underline{r}')(-z^2\delta^{\alpha\beta}
-\sum_{\alpha\beta\kappa\chi}\mathcal{S}^{\alpha\beta\kappa\chi}\nabla_\kappa\nabla_\chi)
-\sum_{\kappa\neq\chi}\nabla_\kappa\Lambda_{ab}^{\alpha\beta\kappa\chi}\nabla_\chi\delta^{\alpha\beta}
\end{align}
\end{subequations}
\end{strip}
By evaluating derivatives of $\langle Z(J)\rangle$ with respect to $J^{\alpha\beta}_{ab}$ at $J^{\alpha\beta}_{ab}=0$, we are able to find the disorder-averaged Green's function of $\Lambda_{ab}^{\alpha\beta\kappa\chi}$.
Integrating $\underline{u}_a$ out in Eq. (48), we obtain a field theory involving only the $\Lambda$ field:
\begin{strip}
\begin{align}
&\langle Z(0)\rangle\propto\lim_{n\rightarrow0}\int\mathcal{D}[\Lambda]\exp\left\{-\frac{1}{2}\sum_{ab=1}^n\sum_{\alpha\beta,\kappa\neq\chi}\left(\ln\det A(\Lambda^{\alpha\beta\kappa\chi}_{ab})+\sum_{\kappa'\neq\chi'}\int d^2rd^2r'\Lambda^{\alpha\kappa\chi\beta}B^{-1}(\underline{r}-\underline{r}')\Lambda^{\alpha\beta\kappa'\chi'}\right)\right\}
\end{align}
\end{strip}
Solving saddle-point problem, we take
\begin{align}
\frac{\delta}{\delta\Sigma^{\alpha\beta}}\left(\text{Tr}\ln A(\Sigma^{\alpha\beta})+\int d^2rd^2r'\Sigma^{\alpha\beta} B^{-1}(\underline{r}-\underline{r}')\Sigma^{\alpha\beta}\right)=0
\end{align}
at $\Sigma_0^{\alpha\beta}$, yielding
\begin{subequations}
\begin{align}
\langle\Sigma^{\alpha\beta}_0\rangle=\frac{1}{2}\sum_{\kappa\neq\chi}\nabla_\kappa B(\underline{r}-\underline{r}')\nabla_\chi \langle G_0(r^\alpha,r'^\beta,z)\rangle\delta^{\alpha\beta}
\end{align}
\begin{strip}
\begin{align}
&G_0(r^x,r'^x,z)=\left[-z^2-(C^{xxxx}+\sigma^2-\sigma^1)\nabla_x\nabla_x-(C^{xxxy}+\Sigma_0^{xx})\nabla_x\nabla_y-(C^{xyxx}+\Sigma_0^{xx})\nabla_y\nabla_x-(C^{xyxy}-\sigma^1-\sigma^2)\nabla_y\nabla_y\right]^{-1}\\
&G_0(r^y,r'^y,z)=\left[-z^2-(C^{yxyy}+\Sigma_0^{yy})\nabla_x\nabla_y-(C^{yxyy}+\Sigma_0^{yy})\nabla_y\nabla_x-(C^{yxyx}+\sigma^2-\sigma^1)\nabla_x\nabla_x-(C^{yyyy}-\sigma^1-\sigma^2)\nabla_y\nabla_y\right]^{-1}\\
&G_0(r^x,r'^y)=\left[-C^{xxyx}\nabla_x\nabla_x-C^{xxyy}\nabla_x\nabla_y-C^{xyyx}\nabla_y\nabla_x-C^{xyyy}\nabla_y\nabla_y\right]^{-1}\\
&G_0(r^y,r'^x)=\left[-C^{yxxx}\nabla_x\nabla_x-C^{yxxy}\nabla_x\nabla_y-C^{yyxx}\nabla_y\nabla_x-C^{yyxy}\nabla_y\nabla_y\right]^{-1}
\end{align}
\end{strip}
\end{subequations}
In $\underline{k}$ space, the condition on the one-particle CPA Green's function may be rewritten as:
\begin{strip}
\begin{subequations}
\begin{align}
&\langle\Sigma^{\alpha\beta}_0\rangle=-\frac{1}{2}\sum_{\kappa\neq\chi}\delta^{\alpha\beta}k_\kappa k_\chi\int d^2q\tilde{B}(\underline{k}-\underline{q})\langle G_0(\underline{q})\rangle;\qquad
\tilde{B}(\underline{k})\equiv\int d^2re^{i\underline{k}\cdot\underline{r}}B(\underline{r})\\
&G_0(k^{x},k^x,z)=\left[-z^2+(C^{xxxx}+\sigma^2-\sigma^1)k_xk_x+(C^{xxxy}+\Sigma_0^{xx})k_xk_y+(C^{xyxx}+\Sigma_0^{xx})k_yk_x+(C^{xyxy}-\sigma^1-\sigma^2)k_yk_y\right]^{-1}\\
&G_0(k^{y},k^y,z)=\left[-z^2+(C^{yxyy}+\Sigma_0^{yy})k_xk_y-(C^{yxyy}+\Sigma_0^{yy})k_yk_x+(C^{yxyx}+\sigma^2-\sigma^1)k_xk_x+(C^{yyyy}-\sigma^1-\sigma^2)k_yk_y\right]^{-1}\\
&G_0(k^{x},k^y)=\left[C^{xxyx}k_xk_x+C^{xxyy}k_xk_y+C^{xyyx}k_yk_x+C^{xyyy}k_yk_y\right]^{-1}\\
&G_0(k^{y},k^x)=\left[C^{yxxx}k_xk_x+C^{yxxy}k_xk_y+C^{yyxx}k_yk_x+C^{yyxy}k_yk_y\right]^{-1}
\end{align}
\end{subequations}
\end{strip}

Applying similar manipulations as for disorder in elastic constants (see Section III), we find that the self-energy scales as
\begin{equation}
\text{Im}\langle\Sigma^{\alpha\beta}_0(k)\rangle\sim-k^2\ln k.
\end{equation}

\section{Steps in the derivation of Eq. (26)}
The Bessel function of the first kind in integral representation is defined as
\begin{equation}
J_n(x)=\frac{1}{\pi}\int_0^\pi\cos(n\theta-x\sin\theta)d\theta.
\end{equation}
When $n=0$, we have
\begin{align}
J_0(x)&=\frac{1}{\pi}\int_0^\pi\cos(x\sin\theta)d\theta=\frac{1}{2\pi}\int_0^{2\pi}\cos(x\sin\theta)d\theta\notag\\
&=\frac{1}{2\pi}\int_0^{2\pi}e^{ix\sin\theta}d\theta=\frac{1}{2\pi}\int_0^{2\pi}e^{ix\cos\theta}d\theta.
\end{align}
We want to calculate
\begin{align}
&\int d^2q\tilde{B}(\underline{k}-\underline{q})\langle G_0(\underline{q},z)\rangle
\propto\int e^{i(\underline{k}-\underline{q})\cdot\underline{r}}\frac{\cos(4\theta)}{r^2+a^2}\langle G_0(\underline{q},z)\rangle d^2qd^2r\notag\\
=&-\int e^{ikr\cos(\theta)-iqr\cos(\phi-\theta)}\frac{r\cos(4\theta)}{r^2+a^2}q\delta(\omega^2-q^2)drdqd\theta d\phi\notag\\
\propto&-\int e^{ikr\cos(\theta)-i\omega r\cos(\phi-\theta)}\frac{r\cos(4\theta)}{r^2+a^2}drdqd\theta d\phi
\end{align}
where $\hat{k}$ is aligned with the x-axis, forming an angle $\phi$ and an angle $\theta$ with $\hat{q}$ and with $\underline{r}$, respectively. On the second line, we have replaced $\langle G_0(\underline{q},z)\rangle$ with $\delta$-function and to obtain the last equality, we used the property of $\delta$-function that $\delta(\omega^2-q^2)=(\delta(\omega-q)+\delta(\omega+q))/(2\omega)$.
We thus can write
\begin{align}
&\int_{-\pi}^\pi\int_{-\pi}^\pi \cos(4\theta)e^{ir(k\cos\theta-\omega\cos(\phi-\theta))}d\theta d\phi\notag\\
=&\int_{\theta=-\pi}^\pi\int_{\tau=-(\theta+\pi)}^{\tau=-(\theta-\pi)}\cos(4\theta)e^{ir(k\cos\theta-\omega\cos\tau)}d\theta d\tau\notag\\
=&\int_{\theta=-\pi}^{\pi}\int_{\tau=-\pi}^\pi \cos(4\theta)e^{ir(k\cos\theta-\omega\cos\tau)}d\theta d\tau\notag\\
=&\int_{\theta=-\pi}^\pi \cos(4\theta)e^{ikr\cos\theta}d\theta\int_{\tau=-\pi}^\pi e^{-i\omega r\cos\tau}d\tau\notag\\
=&\int_{\theta=-\pi}^\pi \cos(4\theta)e^{ikr\cos\theta}d\theta\int_{\tau=-\pi}^\pi e^{i\omega r\cos\tau}d\tau\notag\\
\propto&  J_4(kr)J_0(\omega r)
\end{align}
where we have used the periodicity of trigonometric functions.

\end{appendix}

\balance


\bibliography{rsc} 
\bibliographystyle{rsc} 

\end{document}